\documentclass[letterpaper, 10 pt, conference,twocolumn]{ieeeconf}
\usepackage{amsmath,amsfonts,amssymb}
\usepackage{graphicx,epsfig,psfrag,subfigure,remark}
\usepackage{xfrac,bm}
\usepackage{algorithmic,algorithm}
\usepackage[all]{xy}
\usepackage{varioref}
\usepackage{wrapfig}
\usepackage{threeparttable}
\usepackage{dcolumn}
\newcolumntype{d}{D{.}{.}{-1}}
\usepackage{nomencl}
\makeglossary
\usepackage{subfigure}
\usepackage{comment,cite}
\usepackage{xfrac}
\usepackage{mathrsfs}
\IEEEoverridecommandlockouts 

\newcommand{\x}{{\bm{x}}}
\newcommand{\X}{{\mathsf{X}}}

\newcommand{\z}{{\bm{z}}}

\newcommand{\cE}{\mathcal{E}}

\newcommand{\cG}{\mathcal{G}}

\newcommand{\cI}{\mathcal{I}}

\newcommand{\cP}{\mathcal{P}}

\newcommand{\cS}{\mathcal{S}}

\newcommand{\cT}{\mathcal{T}}

\newcommand{\mT}{T}

\newtheorem{problem}{Problem}

\renewcommand{\t}{^{\mbox{\scriptsize \mT}}}

\renewcommand{\baselinestretch}{0.99}
\newremark{remark}{Remark}

\title{\textbf{Relay Pursuit of an Evader by a Heterogeneous Group of Pursuers\\  Using Potential Games}}

\author{Yoonjae Lee \and Efstathios Bakolas \thanks{This work was supported in part by ARL under W911NF2020085. E. Bakolas (Associate Professor) and Y. Lee (graduate student) are with the Department of Aerospace Engineering
and Engineering Mechanics, The University of Texas at Austin,
Austin, Texas 78712-1221, USA, Emails: bakolas@austin.utexas.edu; yol033@utexas.edu}}

\begin{document}

\maketitle

\begin{abstract}

We propose a decentralized solution for a pursuit-evasion game involving a heterogeneous group of rational (selfish) pursuers and a single evader based on the framework of potential games. In the proposed game, the evader aims to delay (or, if possible, avoid) capture by any of the pursuers whereas each pursuer tries to capture the latter only if this is to his best interest. Our approach resembles in principle the so-called relay pursuit strategy introduced in~\cite{p:bak_auto12}, in which only the pursuer that can capture the evader faster than the others is active. In sharp contrast with the latter approach, the active pursuer herein is not determined by a reactive ad-hoc rule but from the solution of a corresponding potential game. We assume that each pursuer has different capabilities and his decision whether to go after the evader or not is based on the maximization of his individual utility (conditional on the choices and actions of the other pursuers). The pursuers’ utilities depend on both the rewards that they will receive by capturing the evader and the time of capture (cost of capturing the evader) so that a pursuer should only seek capture when the incurred cost is relatively small. The determination of the active pursuer-evader assignments (in other words, which pursuers should be active) is done iteratively by having the pursuers exchange information and updating their own actions by executing a learning algorithm for games known as Spatial Adaptive Play (SAP). We illustrate the performance of our algorithm by means of extensive numerical simulations.

\end{abstract}

\section{Introduction}\label{s:intro}


Pursuit-evasion games (PEGs) with multiple players (also known as group pursuit-evasion games) are receiving a lot of attention at present due to their relevance to applications involving decision makers with possibly conflicting objectives. In a typical multi-player PEGs, there is a group of pursuers playing an adversarial game against another team of agents, which we refer to as the evaders. The goal of the pursuers is to capture the evaders whereas the goal of the evaders is to avoid the capture as long as possible. While many papers have focused on finding reactive rules at the kinematic level that will allow agents to win a game for different problem settings, very little has been said about pursuit-evasion games based on a strategic perspective. In this paper, we present a game-theoretic solution to a special class of PEGs involving multiple pursuers and a single evader. By utilizing the framework of potential games, we show that a meaningful pursuit strategy can naturally emerge as a commonly accepted solution among the pursuers from the same team despite the fact that each pursuer acts selfishly (cooperation emerges from selfish actions).

\textit{Literature review:}  PEGs have been widely studied in various fields, such as defense~\cite{weiss2016minimum,turetsky2003missile}, space~\cite{pontani2009numerical}, and robotics~\cite{chung2011search}. PEGs can be categorized into a few different classes depending on the number of pursuers and evaders participating in the game. In multiple-pursuers single-evader games, for example, some form of cooperation between pursuers is required to capture the evader~\cite{huang2011guaranteed}. In single-pursuer multiple-evaders or multiple-pursuers multiple-evaders games, on the other hand, the pursuer(s) often face an assignment problem~\cite{bakolas2010optimal} or an optimization problem seeking to minimize the costs incurred by the pursuers during the pursuit phase~\cite{fuchs2010cooperative} for assigning themselves to a set of evaders. The numbers of players aside, there could also be scenarios in which the players have restricted sensing abilities, which would require advanced pursuit strategies~\cite{bopardikar2008discrete}. For a broader and more thorough review on recent studies on PEGs, one can refer to~\cite{weintraub2020introduction}.

A particular class of PEGs, which is also the main focus of this paper, is one involving a group of pursuers and a single evader~\cite{p:bak_auto12,p:SUN2017,selvakumar2019feedback}. In such games, although utilizing multiple pursuers to capture the evader is possible, one can instead employ a practical and easy to implement strategy which is known as \textit{relay pursuit}~\cite{p:bak_auto12}. In the latter strategy, only one pursuer goes after the evader (active pursuer) while the rest remain still. The assignment of the evader to one of the pursuers is determined by a simple rule according to which the active pursuer at each instant of time is the one that can capture the evader faster than the other pursuers. In~\cite{p:bak_auto12}, for instance, the active pursuer is assigned by a decentralized (or distributed) algorithm based on dynamic Voronoi diagrams.

A significant portion of the relevant literature has put a lot of emphasis on geometric solution approaches to PEGs including methods based on, for instance, the Apollonius circle~\cite{giovannangeli2010pursuit}, partitions of the boundaries of convex or non-convex objects for border or perimeter defense~\cite{von2020multiple,shishika2020cooperative}, and Voronoi diagrams~\cite{bakolas2010optimal,huang2011guaranteed,p:bak_auto12}, to name but a few. Other approaches rely on combination of optimal control and differential games and numerical techniques for the solution of the Hamilton-Jacobi-Isaacs equation and / or variational inequality ~\cite{fisac2015pursuit,p:tomlin2016}. The majority of the aforementioned approaches give rise to pursuer-evader assignments which are heavily dependent upon the geometry of the configurations of the players and / or their dynamic properties. Such approaches ignore, however, the strategic element of PEGs given that they require that each agent  adopts a strategy which is based on specific (fixed) rules which do not necessarily account for their own individual interests. In this work, we adopt a different approach to multi-player PEGs according to which each player is an autonomous, rational (or selfish) agent who seeks to maximize his own assigned utility function. There have been attempts to tackle a similar problem using a matrix game approach~\cite{selvakumar2017evasion,selvakumar2020min}, yet the pursuer-evader assignment therein is processed in a centralized manner in contrast with our approach which is decentralized.

\textit{Contributions:} In this paper, we address a class of PEGs with multiple pursuers and a single evader by utilizing the framework of potential games. Potential games, which is a special class of non-cooperative games, are guaranteed to admit at least one pure Nash equilibrium. Their defining property is the existence of a potential function which is such that an increase in the (individual) utility of an agent implies an increase in the team's utility. Nash equilibria of potential games can be computed by means of iterative decentralized or even distributed learning algorithms for games~\cite{p:shamma2007}. Some characteristic examples of such algorithms include fictitious play (FP)~\cite{fudenberg1998theory}, regret matching (RM)~\cite{p:shamma2007}, and spatial adaptive play (SAP)~\cite{young2020individual}.

In our work, the individual utility function of every pursuer is defined in accordance with the Wonderful Life Utility~\cite{wolpert2002optimal} which guarantees that the resulting game is a potential game. The team of pursuers will employ the SAP algorithm to compute a pursuer-target assignment profile that corresponds to a pure Nash equilibrium of the corresponding potential game. It turns out that selfish individual actions by the pursuers at each instant of time will eventually lead to a situation in which only one pursuer is active and the others inactive, which is precisely the key idea of the relay pursuit strategy proposed in ~\cite{p:bak_auto12}. We claim that this paper provides a game-theoretic justification of the relay pursuit strategy as well as a way to implement the latter in a decentralized way. Note that the active pursuer or pursuers as determined by our game-theoretic approach do not necessarily match the active pursuer determined by the ad-hoc geometric approach proposed in~\cite{p:bak_auto12}.

\textit{Outline:} The rest of the paper is organized as follows. In Section~\ref{s:PEG_formulation}, we formulate the group pursuit-evasion problem. In Section~\ref{s:game_theoretic_approach}, we analyze the PEG using the framework of potential games.
In Section~\ref{s:siml}, we present and discuss the results of numerical simulations. In Section~\ref{s:concl}, we provide concluding remarks and directions for future research.

\section{Formulation of Pursuit-Evasion Game} \label{s:PEG_formulation}

\subsection{Notations}
Given a vector $\bm{\xi} \in \mathbb{R}^n$, we denote by $\| \bm{\xi} \|$ its Euclidean norm. Let us then denote by $B_\epsilon(\bm{\xi})$ the (closed) ball of radius $\epsilon>0$ centered at $\bm{\xi}$, that is, $B_\epsilon(\bm{\xi}) := \{\bm{x} \in \mathbb{R}^n: \| \bm{x} - \bm{\xi} \| \leq \epsilon \}$. We also denote by $S_\epsilon(\bm{\xi})$ the sphere of radius $\epsilon>0$ centered at point $\bm{\xi}$, that is, $S_\epsilon(\bm{\xi}) := \{\bm{x} \in \mathbb{R}^n: \| \bm{x} - \bm{\xi} \| = \epsilon \}$. Lastly, $[a,b]_d$ denotes the discrete interval from integer $a$ to integer $b$ where $a \leq b$, that is, $[a,b]_d := \{a,a+1,\dots,b\}$.

\subsection{Problem Setup}
We consider a class of pursuit-evasion games involving $N$ pursuers and one evader, all of which are moving in the unbounded 2D plane. We assume that there are no obstacles which hinder the movement of the players. The pursuers will obtain a relevant reward if they successfully capture the evader in time, whereas the evader will continuously strive to avoid or delay the capture. Furthermore, we assume that all players will either move with a constant speed or not move at all. We also assume that all players have a limitless sensing range, meaning that there is no range restriction; the pursuers always know the location of the evader, and vice versa. However, it is worth mentioning here that the previous assumption can be relaxed given that the learning algorithm we will use later on admits a range-constrained (and thus distributed) implementation (we will not discuss the latter in this paper).




The group of pursuers, which is denoted as $\cP :=\{\cP_1,\dots, \cP_N\}$, is assumed to be heterogeneous in the sense that each pursuer has a maximum speed $v_i$, which may differ from one another, for $i\in \cI_{\cP}:=[1,N]_d$ where $\cI_{\cP}$ is the index-set of the group of pursuers $\cP$. We will denote by $\cP_i \in \cP$ the $i^{th}$ pursuer, and by $\cE$ the (single) evader whose maximum speed is denoted as $v_e$. Furthermore, we denote by $\bm{x}_{i} \in \mathbb{R}^2$ (respectively, $\bm{x}_{i}^0 \in \mathbb{R}^2$) the position of $\cP_i$ at time $t\geq 0$ (respectively, $t=0$), where $i \in [1,N]_d$, and by $\bm{x}_{e} \in \mathbb{R}^2$ (respectively, $\bm{x}_{e}^{0} \in \mathbb{R}^2$) the position of $\cE$ at time $t \geq 0$ (respectively, $t=0$). We assume that there is an upper bound $t_f$ on the duration of the game beyond which the pursuers lose the game if the evader has not been captured. Capture of $\cE$ by $\cP_i$ occurs if there is a time instant $t \leq t_f$ at which the latter enters the capture zone of the former, that is, $\bm{x}_{i} \in B_{\epsilon}(\bm{x}_{e}$) for a given capture radius $\epsilon > 0$. The capture time of $\cE$ by $\cP_i$ corresponds to the smallest time $t \leq t_f$ at which capture occurs.

\subsection{Player Dynamics}
 We will assume that all players have single integrator dynamics, that is,
\begin{align}\label{eq:motion}
\dot{\bm{x}}_{i}(t) & = v_{i}\bm{u}_{i}(t), & \bm{x}_{i}(0) &=\bm{x}_{i}^0,\quad i \in [1, N]_d,
\\ 
\dot{\bm{x}}_{e}(t) & = v_{e}\bm{u}_{e}(t), & \bm{x}_{e}(0) &=\bm{x}_{e}^0,
\end{align}
where $\bm{u}_{i} \in \mathcal{U}$ (respectively, $\bm{u}_{e} \in \mathcal{U}$) denotes the control input of $\cP_i$ (respectively, $\cE$) at time $t$, where $\mathcal{U}:= S_1(0) \cup \{0\}$ is the (common) input value set of the players. Note that when $\bm{u}_i \in S_1(0)$ (resp., $\bm{u}_e \in S_1(0)$) then the input corresponds to the direction of motion of $\cP_i$ (resp., $\cE$) whereas when $\bm{u}_i=0$ (resp., $\bm{u}_e = 0$) then $\cP_i$ (resp., $\cE$) does not move.

\subsection{Pure-Pursuit and Pure-Evasion Strategies}
Because the focus of this paper is on the high-level pursuer-evader assignment problem under the assumption that each pursuer is rational, we will prescribe the low-level pursuit and evading strategies. In particular, we will assume that the active pursuers chase the evader by adopting the so-called pure-pursuit strategy, that is,
\begin{equation}\label{eq:pure_pursuit}
\bm{u}_{i} = \bm{\xi}_{i}/\|\bm{\xi}_{i}\|,
\end{equation}
where $\bm{\xi}_{i}:=\bm{x}_{e}-\bm{x}_{i}$ is the relative position vector of $\cE$ with respect to $\cP_i$ under the assumption that $\cE$ views the latter pursuer as the closest pursuer.

The evader, on the other hand, seeks to delay or avoid, if possible, his capture by any of the pursuers. To this aim, we assume that $\cE$ plays a pure evasion strategy against the $k^{th}$ pursuer $\cP_k$ where $k \in \cI_{\cP}$ corresponds to the index of the pursuer who is the closest to $\cE$ and irrespective of whether this pursuer is an active pursuer or not (the pursuers' assignments may not even be known to the evader). The pure evasion strategy is defined as follows:
\begin{equation}\label{eq:pure_evasion}
\bm{u}_{e} = \bm{\xi}_{k}/\|\bm{\xi}_{k}\|,
\end{equation}
where $\bm{\xi}_{k} =\bm{x}_{e}-\bm{x}_{k}$. Note that, when $i=k$, we get $\bm{u}_{e} = \bm{u}_{i}$, where $\bm{u}_{i}$ is defined in \eqref{eq:pure_pursuit} and $\bm{u}_{e}$ in \eqref{eq:pure_evasion}, which means that the pursuer from which the evader tries to flee is the one who is currently chasing him.

\subsection{Time-of-Capture}
Under the assumption that each active pursuer employs the pure-pursuit strategy given in \eqref{eq:pure_pursuit} whereas the evader employs the pure-evasion strategy given in \eqref{eq:pure_evasion}, we can compute (an estimate) of the time it will take for $\cP_i$ to capture $\cE$, or the time of capture denoted by $\phi(\bm{x}_{e_j},\bm{x}_{i}$), by solving the following quadratic equation~\cite{selvakumar2020min}:
\begin{equation}\label{eq:time_metric}
(v_{e_j}^2-v_{i}^2)\phi^2+2(\langle \bm{\xi}_{ij}, v_{e_j}\bm{u}_{e_j} \rangle -\epsilon v_{i})\phi + \|  \bm{\xi}_{ij} \|^2 = \epsilon^2,
\end{equation}
where the evader's (expected) control input $\bm{u}_{e}$ (velocity) is obtained from \eqref{eq:pure_evasion} assuming $\cE$ recognizes $\cP_i$ as the closest pursuer and attempts to avoid from him. Note that we are using the term ``expected'' here for the following reason; if it turns out that $\cP_i$ is not the closest pursuer to the evader (meaning that the underlying assumption about $\cP_i$ being the active pursuer is wrong), then $\cE$ will have a different control input. In such cases, the solution $\phi$ to Equation~\eqref{eq:time_metric} will not correspond to the actual time of capture, but rather an upper bound of the latter.


\section{Game Theoretic Approach}\label{s:game_theoretic_approach}

\subsection{Pursuer-Evader Assignment Problem}
In this section, we will formulate the problem of selecting the active pursuer (pursuer-evader assignment problem) and show that the latter problem can be placed under the umbrella of potential games~\cite{monderer1996potential}, which is a subclass of non-cooperative games. The pursuer-evader assignment problem is, in a nutshell, an optimization problem in which a group of pursuers assign themselves to a group of evaders (a single evader in our case) in a way that their individual utilities can be maximized. Unlike conventional target assignment problems, however, there is one more goal for the pursuers to achieve; that is, they must also aim to minimize the total cost incurred during the process of pursuing the evader(s). The latter objective will be a tacit one given that it is difficult to guarantee an ``efficient'' solution to a game (e.g., efficient Nash equilibrium), especially when one is interesting in computing such a solution by means of a decentralized learning algorithm as we propose to do herein. In this section, we present a way to convert the multiple-evaders single-evader PEG discussed in Section~\ref{s:PEG_formulation} as a pursuer-evader assignment problem by formulating the PEG as a potential game.

\subsection{Non-cooperative Games and Potential Games}
We will briefly review some basic concepts from non-cooperative games and potential games. To this aim, let us denote by $A_i$ the set of available actions of the $i^{th}$ pursuer $\cP_i$, and by $A$ the set of joint actions of the group of pursuers $\cP$, where 
\[
A=  A_1 \times \dots \times A_N.
\] 
We also denote by $U_{\cP_i}:A \rightarrow \mathbb{R}$ the utility function of $\cP_i$; note that $U_{\cP_i}$ is a function of the actions of all the pursuers and not just $\cP_i$. We denote by $a_i$ the action of $\cP_i$; if $a_i = \cE$, then $\cP_i$ is the active pursuer which is assigned to the task of capturing $\cE$. If, on the other hand, $a_i = \varnothing$ then $\cP_i$ is inactive and will stay in the same position (and thus, he will not pursue the evader). Note that we often refer to $\bm{a}$ as $(a_i,a_{-i})$ where $a_i$ is the assignment of $\cP_i$ and $a_{-i}$ represents the joint action profile of the rest of the pursuers. The joint assignment profile of the whole group of pursuers is denoted as $\bm{a}$. Given a non-cooperative game, a joint action profile $\bm{a}^\star$ is a Nash equilibrium, when there is no unilateral motive for each pursuer to choose a different action (assignment) if the other pursuers are committed to their current actions (assignments), that is,
$\forall i \in [1,N]_d$, it holds true that
\begin{equation}\label{eq:noncooperative}
    U_{\cP_i}(a_i^\star,a_{-i}^\star) - U_{\cP_i}(a_i,a_{-i}^\star) \geq 0, \quad \forall a_i\in A_i.
\end{equation}

The distinguishing feature of potential games is the existence of a potential function $\Phi:A \rightarrow \mathbb{R}$ such that $\forall i \in [1,N]_d$ it holds true that
\begin{align}\label{eq:pot}
& U_{\cP_i}(a_i,a_{-i}) - U_{\cP_i}(a'_i,a_{-i})  = \nonumber\\
&~~~~\qquad~~~
\Phi(a_i,a_{-i}) - \Phi(a'_i,a_{-i}), ~~~~\forall a_i \in A_i.
\end{align}
Equation \ref{eq:pot} implies that the difference in the pursuer's utility caused by a change on his own action leads to the same change on the potential function. The existence of a pure Nash equilibrium is guaranteed in potential games. More importantly, one such equilibrium can be computed by means of iterative algorithms known as learning algorithms for games. This is a key property of potential games, which we will leverage in our proposed solution.

\subsection{Utility Design}
We will now define the utilities of the players in such a way that the PEG can be associated with a potential game. An implicit requirement here is that when a pursuer cannot capture the evader within the allotted time, then the former receives negative rewards in case he attempts to capture the latter (his efforts will be in vain).

\noindent \textit{Time-dependent Capture Reward:} We will start by defining the nominal reward function for capturing the evader as follows:
\begin{align}\label{eq:nominal_reward}
r_{\cE}(t)& := t_f-t.
\end{align}
We note that $r_{\cE}(t)$ is an affine function of time $t$ that takes negative values for $t > t_f$. This choice is meant to discourage the pursuers from chasing the evader beyond the final time $t_f$ (if they decide to continue chasing him, they will receive no reward). Note that at $t_0$ (when $t=0$), $r_{\cE}(t_0)=t_f$.

\noindent \textit{Total Time of Capture:} We define the total time of capture to be the sum of the time of capture of the evader by every pursuer who is currently assigned to $\cE$. The total time of capture can be expressed as:
\begin{equation}\label{eq:capture_utility}
T_{\cE}(\bm{a};\bm{x}_e^0,\bm{x}_i^0,t_0):=\sum_{i \in \cI_{\star}(\bm{a};t_0)}\phi(\bm{x}_{e}^0,\bm{x}_{i}^0),
\end{equation}
where $\cI_{\star}(\bm{a};t)$ denotes the index-set of the active pursuers, that is, the pursuers assigned to capture the evader $\cE$, at time $t$.

\noindent \textit{Total Capture Utility:} The capture utility (this is the utility associated to the task of capturing the evader), which is denoted as $U_{\cE}(\bm{a};\bm{x}^0,t)$, is defined to be the difference between the nominal reward and average time of capture, that is,
\begin{align}\label{eq:capture_utility}
U_{\cE}(\bm{a};\bm{x}_e^0,\bm{x}_i^0,t_0) := r_{\cE}(t_0)-\frac{T_{\cE}(\bm{a};\bm{x}_e^0,\bm{x}_i^0,t_0)}{|\cI_{\star}(\bm{a};t_0)|}
\end{align}
where $|\cdot|$ denotes the cardinality of a set. We observe that the utility can be negative if the average time of capture is larger than the nominal reward. This is, again, to encourage the pursuers to remain at the same position rather than following an evaders that they cannot capture in time.

\noindent \textit{Pursuer Utility:} There are many ways to design the pursuer's individual utility function. In particular, we will use the \textit{wonderful life utility} (WLU)~\cite{wolpert2002optimal,p:shamma2007}. The WLU corresponds to the marginal contribution made by the pursuer to the total capture utility, i.e.,
\begin{align}\label{eq:pursuer_utility}
& U_{\cP_i}(\bm{a};\bm{x}_e^0,\bm{x}_i^0,t_0)= U_{\cE}((a_i,a_{-i});\bm{x}_e^0,\bm{x}_i^0,t_0) \nonumber \\
&~~\qquad~\qquad~ - U_{\cE}((a_i=\varnothing,a_{-i});\bm{x}_e^0,\bm{x}_i^0,t_0).
\end{align}

\begin{problem}[Pursuer-Evader Assignment Problem]\label{prob1}
Let us define the initial game of the pursuit-evasion game formulated in Section~\ref{s:PEG_formulation} as the following:
\begin{equation*}
\cG_0 := \langle \cP,\{A_i,U_{\cP_i}(\bm{a};\bm{x}_e^0,\bm{x}_i^0,t_0)\}_{i \in \cI_{\cP}} \rangle.
\end{equation*}
Then, find a
pursuer-evader assignment profile, $\bm{a}_{\star} \in A$, such that $\forall i \in \cI_{\cP}$, the following condition holds true:
\begin{equation*}
U_{\cP_i}((a_{i}^{\star},a_{-i}^{\star});\bm{x}_e^0,\bm{x}_i^0,t_0) \geq U_{\cP_i}((a_{i},a_{-i}^{\star});\bm{x}_e^0,\bm{x}_i^0,t_0)
\end{equation*}
$\forall a_{i} \in A_{i}$, which corresponds to a pure Nash equilibrium of the game $\cG_0$.
\end{problem}

Finally, the WLU guarantees the existence of a potential function ~\cite{wolpert2002optimal}, which, in our case, turns out to be the total capture utility $U_{\cE}$.
In particular, by setting $\Phi=U_{\cE}$ we get:
\begin{align*}
& U_{\cP_i}((a'_i,a_{-i});\bm{x}_e^0,\bm{x}_i^0,t_0) - U_{\cP_i}((a_i,a_{-i});\bm{x}_e^0,\bm{x}_i^0,t_0) \nonumber \\
& = \left( U_{\cE}((a'_i,a_{-i});\bm{x}_e^0,\bm{x}_i^0,t_0) - U_{\cE}((\varnothing,a_{-i});\bm{x}_e^0,\bm{x}_i^0,t_0) \right) \nonumber \\
& ~~ - \left( U_{\cE}((a_i,a_{-i});\bm{x}_e^0,\bm{x}_i^0,t_0) - 
U_{\cE}((\varnothing,a_{-i});\bm{x}_e^0,\bm{x}_i^0,t_0) \right) \nonumber \\
& = U_{\cE}((a'_i,a_{-i});\bm{x}_e^0,\bm{x}_i^0,t_0) - U_{\cE}((a_i,a_{-i});\bm{x}_e^0,\bm{x}_i^0,t_0), \nonumber
\end{align*}
from which we conclude that
\begin{align}\label{eq:potential}
& U_{\cP_i}((a'_i, a_{-i}); \bm{x}_e^0,\bm{x}_i^0,t) - U_{\cP_i}((a_i, a_{-i}); \bm{x}_e^0,\bm{x}_i^0,t) \nonumber \\
& ~ = \Phi((a'_i, a_{-i});\bm{x}_e^0,\bm{x}_i^0,t) - \Phi((a_i, a_{-i});\bm{x}_e^0,\bm{x}_i^0,t_0).
\end{align}
Equation~\eqref{eq:potential} implies that $\cG_0$ is an exact potential game.


\subsection{Spatial Adaptive Play}

Given the fact that our PEG can be formulated as a potential game, we can now compute a pure Nash equilibrium by utilizing any iterative learning algorithm for potential games. The algorithm we will utilize, SAP, is one that is guaranteed to converge to a nearly efficient pure Nash equilibrium
which yields almost but not the highest total capture utility~\cite{p:shamma2007}. In particular, SAP is suitable for multi-agent target assignment problems with relatively a small number of players whereas for a large number of players, one can utilize instead the Selective Spatial Adaptive Play algorithm~\cite{p:shamma2007}. 

We will denote by $\psi$ the update law (based on SAP) of the evader assignment profile of the pursuers. The pseudocode for SAP is provided in Algorithm~\ref{alg:sap}, whose main steps are discussed next. At every simulation step, a random pursuer $\cP_i$ is selected. Thereafter, $\cP_i$ computes all the hypothetical utilities he could obtain by selecting every action in his current action profile $A_i$ and store the results in a set $\cS_i$ (Line~\ref{sap:line:3}). The goal of Line~\ref{sap:line:5} is to obtain probability distribution over the action profile $A_i$. Based on this probability distribution, the pursuer will choose his next action (Line~\ref{sap:line:6}). This can be achieved by using the following equation:
\begin{equation}\label{eq:softmax}
\sigma(\bm{x})=\frac{e^{\tau^{-1} \bm{x}}}{\sum_{k=1}^{|\bm{x}|} e^{\tau^{-1} \bm{x}_k}},
\end{equation}
where $\tau$ is a parameter for the randomization level that prevents the pursuer from stagnating in local maxima during the early phase of the game. 

\begin{algorithm}
 \caption{Spatial Adaptive Play: $\psi$}
 \begin{algorithmic}[1]
 \renewcommand{\algorithmicrequire}{\textbf{Input:} $A_i, a_{-i}, \bm{x}_e, \bm{x}_i, t$}
 \renewcommand{\algorithmicensure}{\textbf{Output:} $a_i$}
 \REQUIRE 
 \ENSURE  
  \STATE {$\cS_i \leftarrow \varnothing$}
  \FOR {$a'_i \in A_i$}
  \STATE {$\cS_i = \cS_i \cup \{ U_{\cP_i}((a'_i,a_{-i});\bm{x}_e,\bm{x}_i,t)\}$}
  \label{sap:line:3}
  \ENDFOR
  \STATE {$p_i = \sigma(\cS_i)$} \label{sap:line:5}
  \STATE {$a_i \leftarrow$~\texttt{RandSample}($A_i;p_i$)}
  \label{sap:line:6}
  \RETURN $a_i$
 \end{algorithmic} 
 \label{alg:sap}
 \end{algorithm}

\subsection{Dynamic Evader Assignment}

When game-theoretic learning algorithms, such as SAP, are used to solve potential games, there is an implicit assumption that the game is static, meaning that the states and utilities of the players are time-invariant; with this assumption it is guaranteed that the learning algorithm will converge to a Nash equilibrium. The stationary assumption is, however, not practical in our pursuit-evasion game for a number of reasons. First, it is practically infeasible for the agents to run hundreds or thousands of negotiation protocols fast enough until the evader moves to a different location. Second, the evader's state and utility rapidly change with time in continuous time, and this change will not be reflected in the negotiations if the game is treated as a static one. Third, even if the pursuers successfully agree upon an optimal assignment for game $\cG_0$ at time $t=0$, this assignment may no longer be the optimal solution for the following games $\cG_t$ for $t>0$ where $\cG_t := \langle \cP,\{A_i(t),U_{\cP_i}(\bm{a}(t);\bm{x}_e(t),\bm{x}_i(t),t)\}_{i \in \cI_{\cP}} \rangle$.
One way to overcome this issue is, albeit not wise, to stop the clock and let the pursuers re-negotiate from the beginning to update their joint assignment periodically. This method, however, will obviously cause a significant amount of computation burden.

Instead, we will utilize the so-called dynamic evader assignment method~\cite{bakolas2020decentralized}, whose main steps are summarized in Algorithm~\ref{alg:dynamic_assignment}. Herein, the negotiation process takes place on-the-fly while the pursuers conduct their most current assignment. In Line~\ref{dea:line:4}, every pursuer selects a random action at $t=0$ (either follow or stay since there is only one evader in this game). Thereafter, at every time step, one pursuer is randomly selected (Line~\ref{dea:line:7}) and updates his action using the SAP update law (Line~\ref{dea:line:10}), while the others continue to conduct their past assignments (Line~\ref{dea:line:11}). By utilizing this method, we will be able to not only reduce the computational burden of each pursuer but also enable the pursuers to converge to a near-optimal assignment $\bm{a}(t)$ for $\cG_t$.

\begin{algorithm}
\caption{SAP-based dynamic evader assignment}
\begin{algorithmic}[1]
\renewcommand{\algorithmicrequire}{\textbf{Input: $\cP, \cE, \psi, t_f, \delta t$}}
\renewcommand{\algorithmicensure}{\textbf{Output:} Dynamic pursuer-evader assignment}
\REQUIRE 
\ENSURE  
\WHILE {$t \leq t_f$}
\IF {$t=0$}
\FOR {$i \in \cP$}
\STATE {$A_i(t_0) \leftarrow$ Update action profile}
\STATE {$a_i(t_0) \leftarrow$
\texttt{RandSample}($A_i(t_0)$)}
\label{dea:line:4}
\ENDFOR
\ELSE
\STATE {$i \leftarrow$
\texttt{RandSample}($\cP$)}
\label{dea:line:7}
\STATE {$\bm{x}_e(t),\bm{x}_i(t) \leftarrow$ Update player states}
\STATE {$A_i(t) \leftarrow$ Update action profile}
\STATE {$a_{i}(t)=\psi(A_i(t),a_{-i}(t-\delta t);\bm{x}_e(t), \bm{x}_i(t),t)$}
\label{dea:line:10}
\STATE $a_{-i}(t) \leftarrow  a_{-i}(t-\delta t)$
\label{dea:line:11}
\ENDIF
\ENDWHILE
\end{algorithmic} 
\label{alg:dynamic_assignment}
\end{algorithm}

\section{Numerical Simulations}\label{s:siml}

In this section, we present numerical simulation results obtained with the application of the SAP-based dynamic evader assignment algorithm (Algorithm~\ref{alg:dynamic_assignment}) for the solution of the pursuit-evasion game (Problem~\ref{prob1}) and compare these results with those obtained by executing Algorithm~\ref{alg:centralized_solution} which yields a benchmark solution to our problem. Algorithm~\ref{alg:centralized_solution} is a centralized ad-hoc algorithm which solves the pursuer-evader assignment problem in the following way: on behalf of the individual pursuers, a central server will, by updating the active pursuer-evader assignment at each time step until the evader is captured or the final time $t_f$ has elapsed, collect the time metric of every pursuer (Line~\ref{cea:line:alg3line5}), select the pursuer with the smallest time metric (Line~\ref{cea:line:alg3line7}), and order him to chase the evader while the other pursuers remain still (Line~\ref{cea:line:alg3line9} - Line~\ref{cea:line:alg3line10}). To that end, the central server must know the state of every pursuer as well as the state of the evader. 

\begin{algorithm}
 \caption{Centralized pursuer-evader assignment}
 \begin{algorithmic}[1]
 \renewcommand{\algorithmicrequire}{\textbf{Input:} $\cP,\cE,\phi,t,t_f$}
 \renewcommand{\algorithmicensure}{\textbf{Output:} $\bm{a}(t)$}
 \REQUIRE 
 \ENSURE
  \STATE {$\bm{x}_e(t),\bm{x}_i(t) \leftarrow$ Update player states}
  \STATE {$\cT \leftarrow \varnothing$}
  \FOR {$i \in \cI_{\cP}$}
  \STATE {$\cT=\cT \cup \{\phi(\bm{x}_e(t),\bm{x}_i(t))\}$}
  \label{cea:line:alg3line5}
  \ENDFOR
  \STATE {$i \leftarrow \arg\min \cT$}
  \label{cea:line:alg3line7}
  \IF {$\cT(i) \leq t_f$}
  \STATE {$a_i(t) \leftarrow \cE$}
  \label{cea:line:alg3line9}
  \STATE {$a_{-i}(t) \leftarrow \varnothing$}
  \label{cea:line:alg3line10}
  \ENDIF
 \end{algorithmic} 
 \label{alg:centralized_solution}
 \end{algorithm}

In our simulations, we assume that there always exists at least one pursuer whose maximum speed is greater than that of the evader, which we refer to as the \textit{super pursuer}. We assume the existence of the super pursuer in order to guarantee that the team of pursuers committed to employ the pure pursuit strategy will have good chances to capture the evader at some time $t \leq t_f$. As we have mentioned, the focus of the paper is on the high-level pursuer-target assignment problem (Problem~\ref{prob1}) rather than the low-level pursuit problem. Indeed, capture of a faster evader may still be possible in the latter case but our interest here is to generate a significant number of scenarios with successful capture of the evader by at least one pursuer in order to assess the proposed game-theoretic solution in contrast with the centralized execution of the relay pursuit strategy proposed in \cite{p:bak_auto12}, in which an assumption that ensures that the evader will always be captured by the pursuer is also made. Furthermore, since the game is simulated in discrete time (although the equations of motion of the pursuers are described in continuous time), we will play a turn-based simulation in which the evader takes an action (evasion) first and then the pursuers will take actions accordingly. If the actions of any pursuer lead him to enter the capture zone of the evader at time $t\leq t_f$, then the game will be terminated (successful capture of the evader).

\begin{figure}[htbp]
\centerline{\includegraphics[scale=0.4]{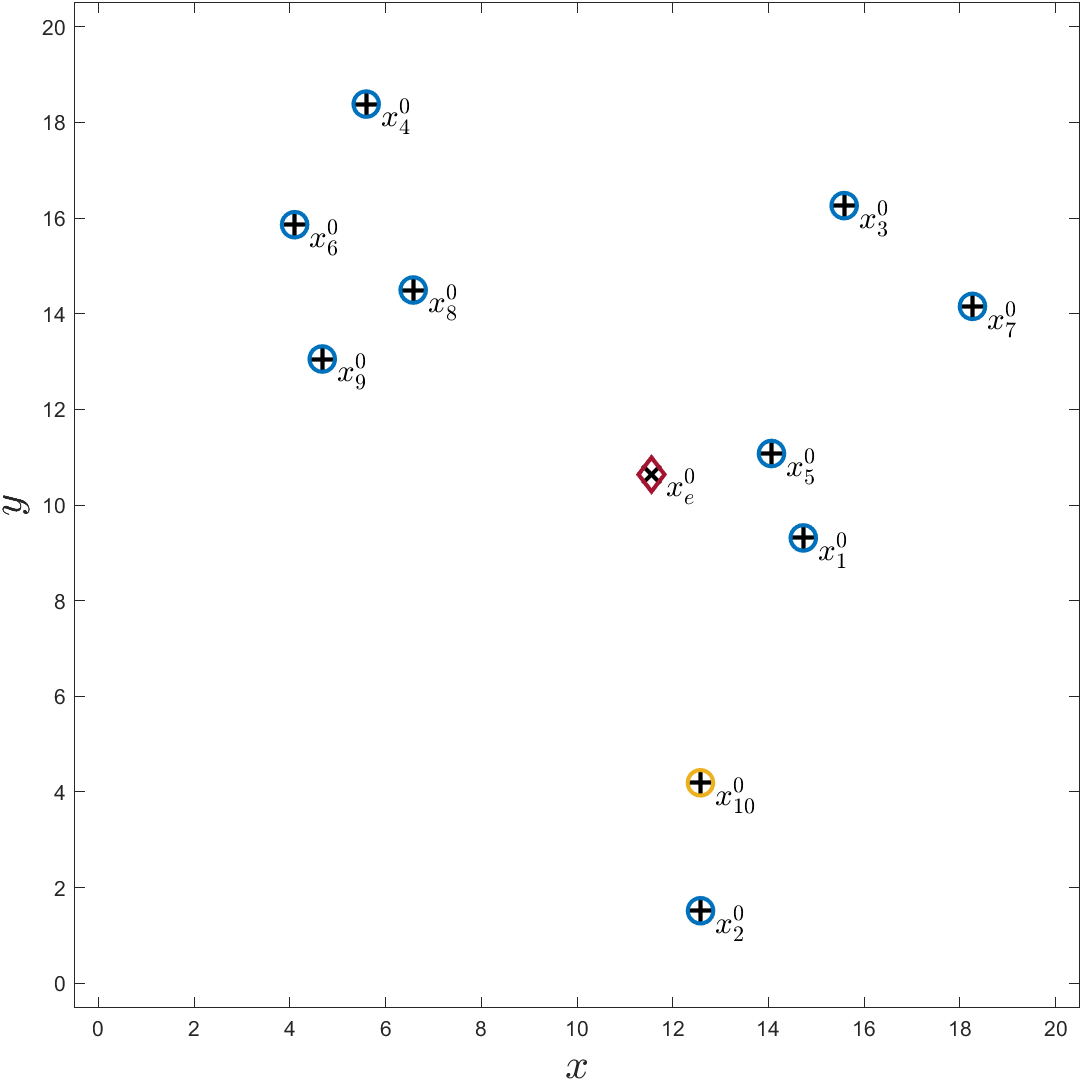}}
\caption{Initial configuration of the PEG ($N=10$)}
\label{fig:PEG_initial_config}
\end{figure}

\begin{figure}[htbp]
\centering
\subfigure[Centralized, $\delta t=0.1$]{\includegraphics[width=4.2cm]{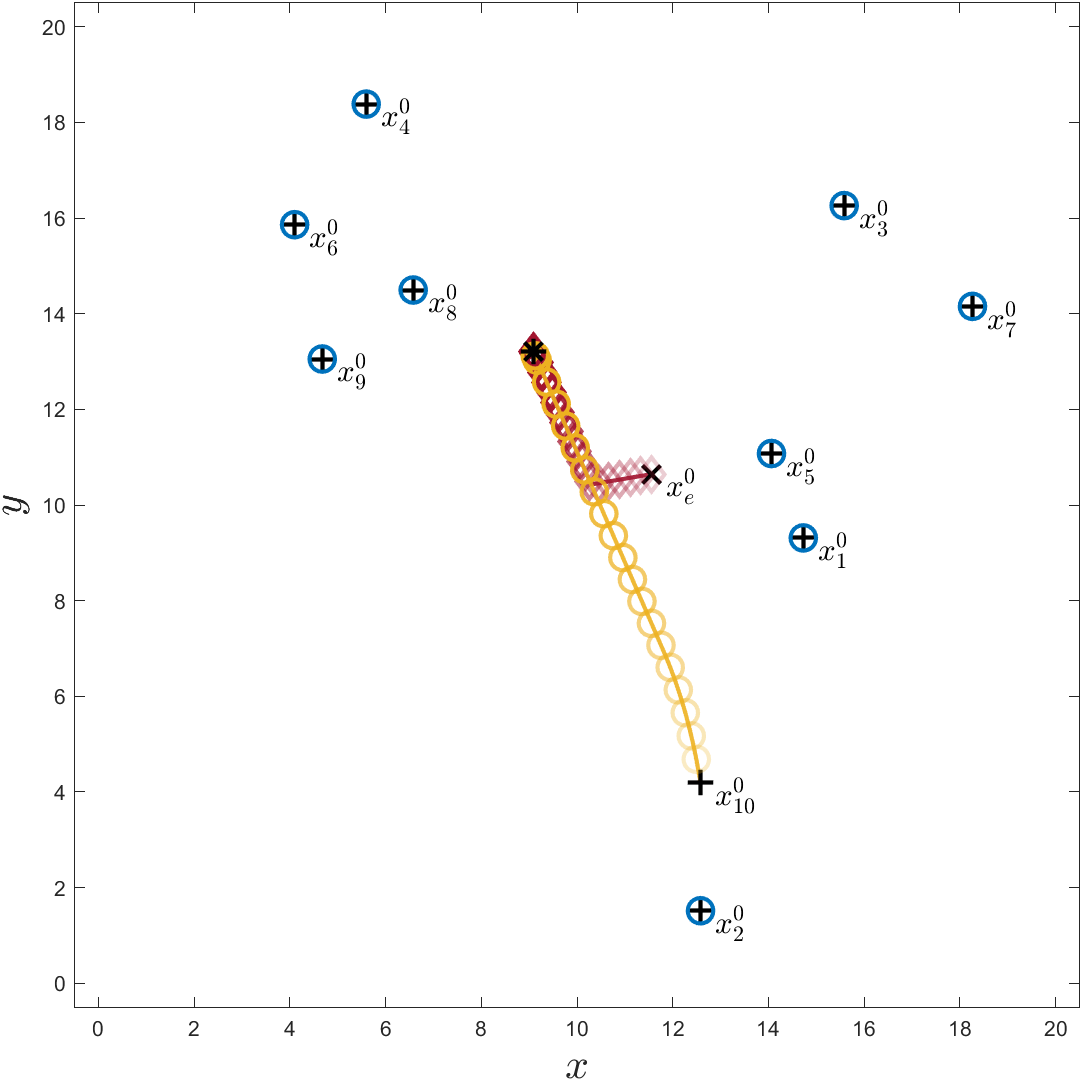}\label{fig:traj_S0}}
\subfigure[Decentralized, $\delta t=0.1$]{\includegraphics[width=4.2cm]{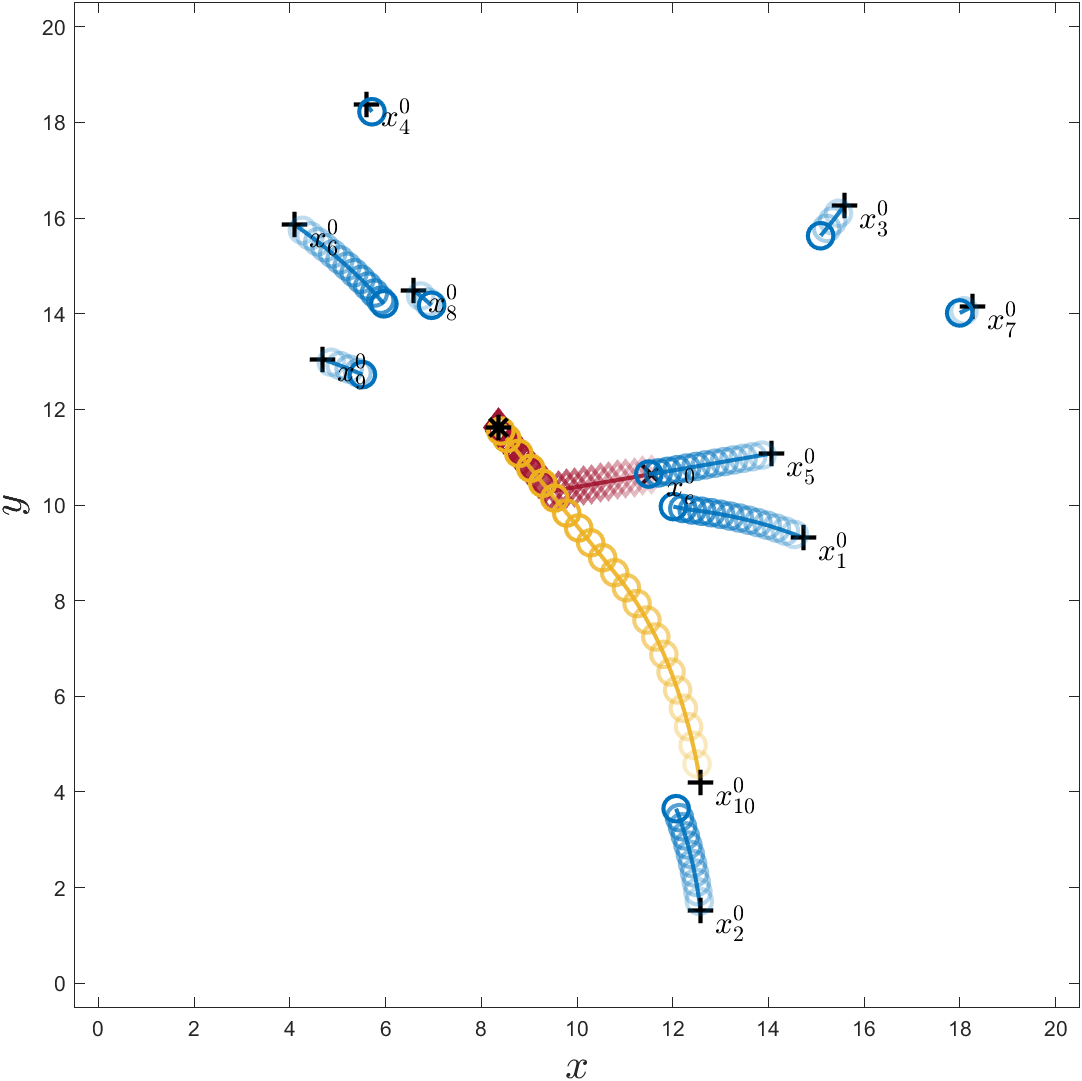}\label{fig:traj_S1}}
\subfigure[Decentralized, $\delta t=0.01$]{\includegraphics[width=4.2cm]{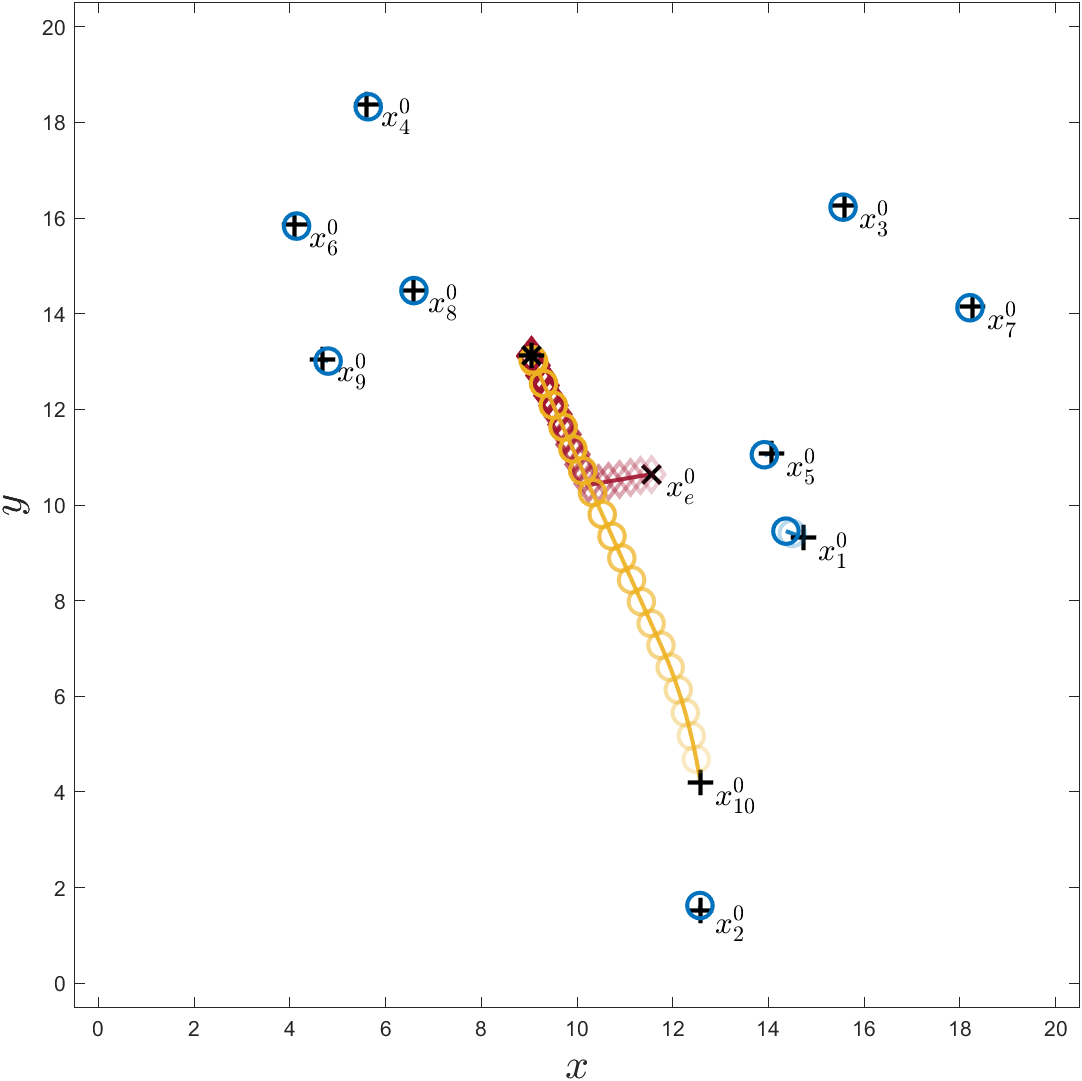}\label{fig:traj_S2}}
\subfigure[Decentralized, $\delta t=0.001$]{\includegraphics[width=4.2cm]{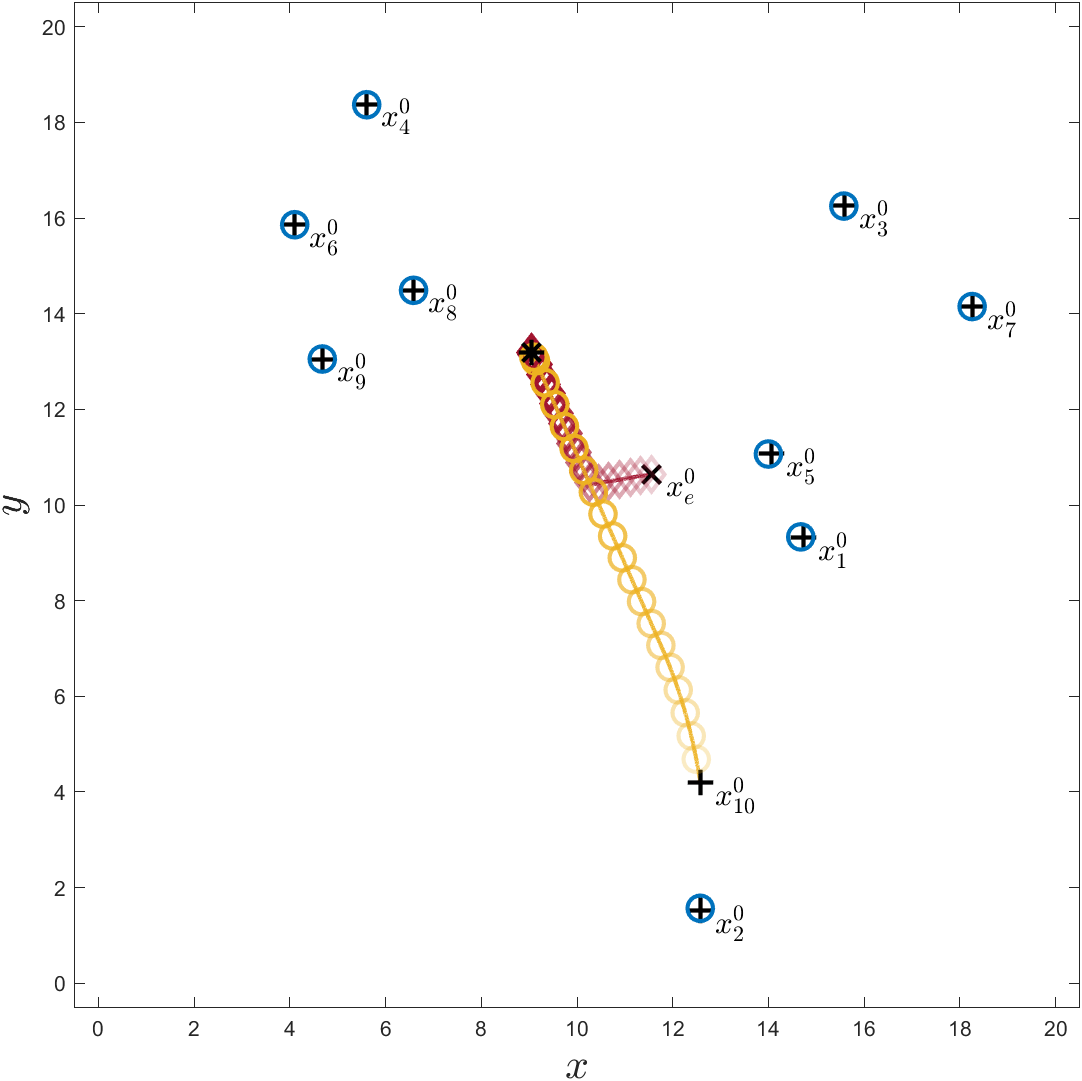}\label{fig:traj_S3}}
\caption{Pursuit-evasion trajectories of pursuers and evaders}
\label{fig:PEG_traj}
\end{figure}

\begin{figure*}
\centering
\subfigure[Decentralized, $\delta t=0.1$]{\includegraphics[width=5.8cm]{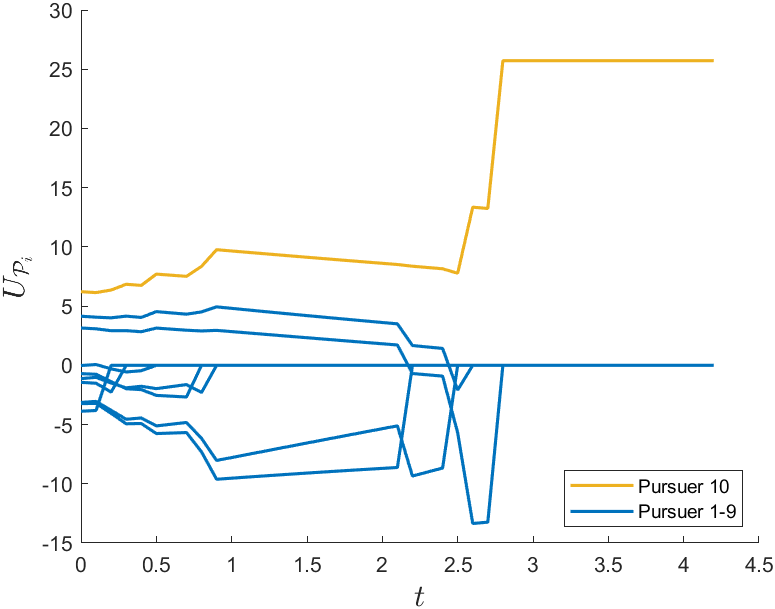}\label{fig:util_S1}}
\subfigure[Decentralized, $\delta t=0.01$]{\includegraphics[width=5.8cm]{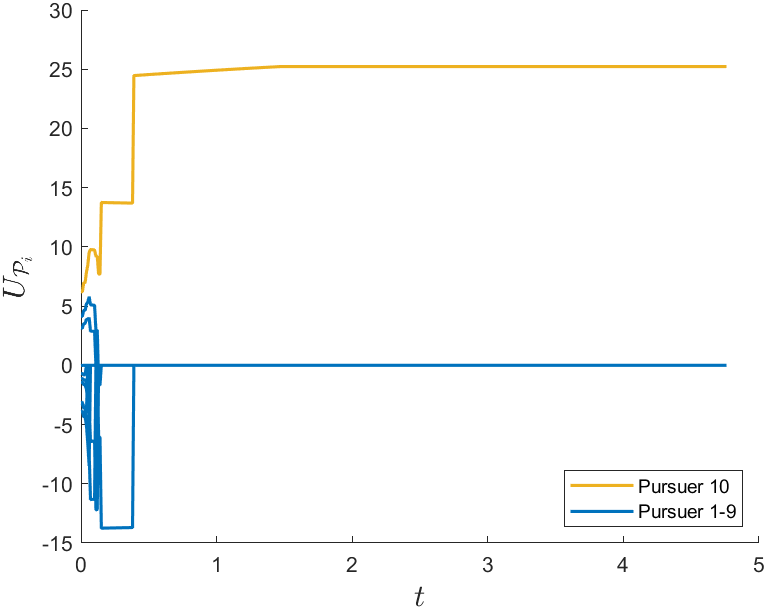}\label{fig:util_S2}}
\subfigure[Decentralized, $\delta t=0.001$]{\includegraphics[width=5.8cm]{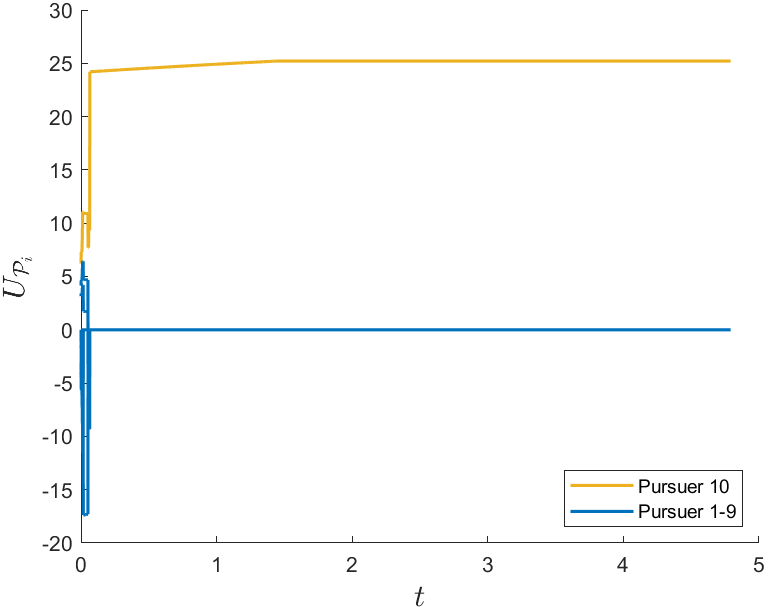}\label{fig:util_S3}}
\caption{Individual utilities of pursuers versus time}
\label{fig:PEG_utilities}
\end{figure*}

On top of the scenario using the centralized solution Algorithm~\ref{alg:centralized_solution}, we will consider three additional scenarios which are based on our decentralized solution with sampling period $\delta t \in \{0.1, 0.01, 0.001\}$. These scenarios allow us to examine how the choice of the sampling period affects the performance of our SAP-based dynamic assignment algorithm during the game. The common parameters used in all of the three decentralized scenarios as well as the centralized scenario are: $N=10,t_f=30,x_i^0 \in [0,20],x_e^0 \in [8,12],v_i=1\ \forall i \in [1,9]_d,v_{10}=2,v_e=0.9,\epsilon=0.1$, and $\tau=10/\eta^2$ where $\eta=t/\delta t$. In summary, the game comprises of 10 pursuers and one evader, and the $10^{th}$ pursuer $\cP_{10}$ is the super pursuer (as well as the fastest player in the game) whose maximum speed $v_{10} = 2$, whereas $v_i = 1$ for $i \in [1,9]_d$. Furthermore, the maximum speed of the evader $v_e=0.9$. The initial configuration of the players is shown in Figure~\ref{fig:PEG_initial_config}; therein, $\cP_{10}$ is depicted as a yellow circle, the other pursuers as blue circles, and the evader as a red diamond.

Figure~\ref{fig:PEG_traj} clearly shows that, for the simulations based on Algorithm~\ref{alg:dynamic_assignment}, the smaller $\delta t$ we pick, the more similar the pursuers' trajectories will look like with those obtained with the execution of the centralized approach. In other words, the pursuers tend to adapt to the relay pursuit strategy more quickly with small $\delta t$. This is demonstrated in Figure~\ref{fig:traj_S1} where we can compare the speed of convergence in each scenario. Conversely, if convergence is too slow, many pursuers will stay active in the early phase and end up wasting their resources (such as fuel or energy). Table~\ref{tab:1} shows the total sum of the time during which each pursuer was active (all the numerical values in this table are averaged data from $10^2$ simulation runs). Lastly, note that, although $\cP_{10}$ is the super pursuer herein, he is not the only pursuer who captures the evader; we occasionally observe in some simulations that a pursuer different from $\cP_{10}$ who is located in the vicinity of the evader captures him while $\cP_{10}$ remains still.

All the simulation data presented herein show that, for a class of PEGs involving multiple pursuers and a single evader, individually selfish action by the pursuers guided by a game-theoretic learning algorithm (negotiation protocol) naturally converge to a joint strategy which is equivalent to the relay pursuit strategy (the latter is executed in an ad-hoc centralized way that yields the benchmark solution to the game). Therein, despite acting selfishly, the pursuers end up yielding the opportunity of capturing the evader to the pursuer that can capture the evader with the shortest amount of time so that the pursuers as a team can obtain the maximized capture utility while minimizing the total time of capture.

\begin{table}[htbp]
\caption{Sum of individual pursuer's active time until capture of evader ($t_f=30$)}
\begin{center}
\begin{tabular}{|c|c|c|c|c|}
\hline
\multicolumn{1}{|c|}{}&\multicolumn{2}{c|}{Centralized}&\multicolumn{2}{c|}{Decentralized} \\
\cline{1-5} 
$\delta t$ & \textit{$N=20$}& \textit{$N=40$}& \textit{$N=20$}& \textit{$N=40$} \\
\hline
0.1& 4.6000& 4.8000& 55.5900& 116.4590\\
\hline
0.01& 4.5700& 4.6900& 11.6580& 39.0082\\
\hline
0.001& 4.5580& 4.6700& 5.2088& 7.9598\\
\hline
\end{tabular}
\label{tab:1}
\end{center}
\end{table}

\section{Conclusions}\label{s:concl}

In this paper, we have presented a decentralized approach to the pursuer-evader assignment problem that arises in pursuit-evasion games involving multiple pursuers and a single evader. In our approach, which relies on non-cooperative game theory, all the pursuers are assumed to be rational decision makers who try to maximize their own utilities (conditional on decisions of their teammates). By designing the utilities such that the pursuer-evader game can put under the umbrella of potential games, some interesting pursuers' behaviors emerge. In particular, the selfish pursuers end up agreeing upon a relay pursuit strategy by executing a decentralized learning algorithm known as spatial adaptive play. In our future work, we will consider an extension of our proposed approach to multiple-pursuer multiple-evader PEGs. We will also consider problems with incomplete information in which the evader can adopt more advanced and sophisticated evasion strategies.

\bibliographystyle{ieeetr}
\bibliography{bakolas,pegref}
\end{document}